# Periodically Spaced CaF₂ Semi-Insulating Thin Ribbons Growth Study on the Si(100) Surface

Eric Duverger[1], Damien Riedel[2*]

[1]Institut FEMTO-ST, Univ. Bourgogne Franche-Comté, CNRS, 15B avenue des Montboucons, F-25030 Besançon, France.

[2]Institut des Sciences Moléculaires d'Orsay (ISMO), CNRS, Univ. Paris Sud, Université Paris-Saclay, F-91405 Orsay, France.

**Abstract :**

The use and the study of semi-insulating layers on metals and semiconductors surfaces have found continuous interest in the past decades. So far, the control of the sizes and growth location of the insulating islands on the substrate is either ill-defined or usually constrained to the use of evaporation masks which size can easily exceed tenth of nanometers. Here, we show that it is possible to grow self-organized periodically spaced thin ribbons of semi-insulating stripes on the bare Si(100) surface. The epitaxial growth of these structures is obtained by the evaporation of CaF₂ molecules on the silicon surface with a coverage of 1.2 monolayers. They are investigated via scanning tunneling techniques at low temperature (9K). The obtained ribbons exhibit a surface bandgap of ~3.2 eV as well as a resonant state at the central part of the ribbons at ~2.0 eV below the Fermi level energy. The use of the density functional theory allows suggesting a model structure of the observed ribbons and reproducing the experimental STM topographies. The formation of the thin ribbons is discussed and we point out the influence of the mechanical forces inside and between the structures that may influence their periodicity.

* Corresponding author: damien.riedel@universite-paris-saclay.fr

# 1- INTRODUCTION

It is now well established that various semi-insulating layers can be grown on metal surfaces such as $Al_2O_3$ or NaCl[1,2]. On semiconductor surfaces, the growth of material having similar insulating properties on the Si(111) surface is very well documented and has been demonstrated with the use of various elements such as $SiO_2$, $SrF_2$, or $CaF_2$[3,4,5,6]. The use of the Si(100) surface for the same purpose has only been explored in the past decade through the epitaxy of $CaF_2$ molecules[7,8,9]. At the first stage, the work of Pasquali et al. were pioneer when using photoemission techniques to describe the mesoscopic aspects of the epitaxy[12]. Consequently, by exploiting local probe techniques at low temperature, it has been very recently shown that thin semi-insulating islands of $Ca_xF_y$ can form large (> 40 nm) ribbons of stripes (LRS) providing its detailed structure at the atomic scale[7]. While at relatively high coverage, the Stanski-Krastanov (SK) process governs most of these epitaxial growths via the formation of an initial wetting layer, the final 2D structures obtained on the surface generally form homogeneous islands of various sizes, shapes and thicknesses[10,11]. The internal structure of these islands usually follows the crystallographic properties of the supporting surface especially because, for the present case, the lattice mismatch between $CaF_2$ ($a_0 = 5.451$ Å) and silicon ($a_0 = 5.43$ Å) is relatively small [7, 8 ,9 ,12]. Yet, some structural modulations between the surface and the insulating layer have already been observed, which is mainly due to the formation of the WL that partly etches the Si(100) surface [8, 13]. However, the localization of the insulating islands on the surface, at the initial epitaxial stage, is mostly random as it is generally ruled by the formation of nucleation sites. The use of vicinal surfaces can bring to a partial control of the localization of the epitaxial process, yet, without being on the same terrace[5]. Henceforth, monitoring the positioning as well as the distance separation (i.e. the periodicity) of such insulating islands on the same terrace can be of crucial importance to exploit their properties at the nanoscale and for their interconnection with other structures[14,15,16]. Our article aims to describe and explain the formation of a new periodically spaced parallel thin ribbons structure and to identify this new 2D assembly observed when $CaF_2$ is exposed to a bare Si(100) surface at a slightly higher temperature than that used in previous work. Our investigations describe the periodicity of the observed structure as well as its electronic properties, which exhibit electronic confinement effects along the central axis of their structure. A theoretical investigation using the density functional theory allows us to suggest a surface structure that can reproduce our experimental observations. This work opens the possibility to periodically localize thin ribbons of stripes that present similar insulating properties than the recently studied LRS.

## 2 – EXPERIMENTAL AND THEORETICAL SECTION

### 2.1 Surface preparation

The experimental measurements presented in this article are made with a low temperature scanning tunneling microscope (STM equipped with a PanScan head from CREATEC) running at 9 K. The studied surfaces are initially prepared with a bare Si(100) surface using n-type doped silicon samples wafers (As doped, $\rho$ = 5 $\Omega$.cm). The Si(100) samples used are primarily cleaned from their initial oxide layer and then followed by their surface reconstruction. These procedures are very well described in the literature and also reported in various previous work[17,18].

Following this preparation procedure, the surface temperature is then stabilized at 740 °C and placed in front of an effusion cell in which bulk $CaF_2$ crystal pieces are heated up to 950 °C to reach the sublimation of $CaF_2$ molecules. Note that the formation of the previously studied LRS structure is obtained when the silicon surface is kept at 710 °C. The quartz balance is formerly calibrated to reach an evaporation rate ~ 0.1 - 0.2 ML.mn$^{-1}$. This evaporation rate is kept relatively low compared to previous work to favor soft landing of the $CaF_2$ molecules on the heated silicon surface. The pristine silicon surface is thus exposed to the $CaF_2$ molecules flux to obtain a coverage rate of ~ 1.2 – 1.3 monolayer (ML). The sample holder is then slowly cooled down to ~12K to be transferred inside the STM scanner. During the surface analysis, the STM topographies are obtained at bias voltages applied on the silicon surface in the range -4 V to +4 V, with tunnel current ~ 5 - 50 pA. The dI/dV spectroscopy curves as well as the dI/dV mapping of the structures are obtained by using a lock-in amplifier running with a 10 mV amplitude modulation at a frequency of 853 Hz.

## 2.2 Theoretical methods

The growth of $CaF_2$ in the experimental conditions is carried out theoretically, via the use of the density functional theory (DFT) as implemented in the Spanish Initiative for Electronic Simulations with Thousands of Atoms (SIESTA) code[19,20]. The calculations are performed using a polarized double-ζ basis set (DZP) with Perdew–Burke–Ernzerhof (PBE) functional in the generalized gradient approximation (GGA) scheme for exchange correlation[21]. The LMKLL version of the non-local van der Waals potential is included to describe the exchange-correlation[22]. A mesh cutoff of 150 Ry with 16 k-points for the Brillouin zone integration is considered to calculate the total energies within a numerical precision of 1 meV. In a first step, in order to reproduce a single thin ribbon unit cell, the geometry relaxation of a $CaF_2$/Si(100) slab containing 4 $CaF_2$ stripes (i.e. $5.78 \times 1.55 \times 4.44$ nm$^3$) of various orientations is performed by the conjugate-gradient method with a force convergence criterion of 0.02 eV/Å. In a second step, the previously optimized slab is doubled in the [1-10] direction (i.e. $11.56 \times 1.55 \times 4.44$ nm$^3$) to reconstruct the $CaF_2$/Si(100) structure containing 8 $CaF_2$ stripes. We then proceed to the annealing and relaxation of the whole system with the same convergence criterion previously used. Following these procedures, various single point energy calculations can be performed such as the localized density of state (LDOS) of the second slab structure to compare it with experimental results. To simulate the experimental annealing and reproduce the dynamics of the growth formation of the thin ribbons, the random initial velocities, corresponding to a temperature of 1000 K are assigned to all the atoms of the surface by using Maxwell–Boltzmann distribution. As described above, the energy minimization is performed on the whole system via the relaxation of the slab forces to obtain the minimum energy of the entire slab. However, some residual atomic forces remain on each Ca atom due to local electrostatic interactions. These forces can be computed and plotted over each axis to describe the specific interactions inside and between the thin ribbons of stripes.

## 3 RESULTS AND DISCUSSIONS

Following the evaporation of $CaF_2$ on the Si(100) surface as described in section 2.1, the resulting observed surface is shown in Fig. 1a. One can clearly observe a succession of thin ribbons equally spaced on the main left terrace. The same periodic structures are not observed on the right part of the STM image where a succession of small terraces is formed. On these terraces, and on the left side of the area where periodic thin ribbons are formed, one can recognize the typical structure of a wetting layer (WL) (Fig. 1a)[7,8,9]. A detailed aspect of the periodic thin ribbons is presented in Figs. 1b and 1c. In particular, in Fig. 1c, it is clearly possible to distinguish the fine internal structure of each ribbon made of two thin light gray lines sandwiched by two sided edges of brighter intensity. The STM images in Fig. 1 are acquired in constant current mode and, hence, the brightness changes traduce a variation of STM tip height, i.e. a variation of density of state at the chosen bias.

From these variations of tip height, the ensuing measured profile across two ribbons (red line in Fig. 1c) is depicted in Fig. 1d. The sizes of each ribbon are very similar, having a width of ~ 26 Å from two maxima. The inner structure of each thin ribbon reveals two central thin stripes separated by 4.5 Å while the period between ribbons is ~ 55 Å. Additionally, one can notice that each thin ribbon is periodically spaced out by 30 Å. As a comparison, previous work reports that large $CaF_2$ ribbons of stripes (i.e. LRS) grown at a lower temperature (710 °C) are made of ribbons of stripes having of single periodicity of ~ 11.7[7].

The electronic properties of the periodic ribbons can be described via inelastic electron tunnel spectroscopy (IETS) technique at specific points on the structure. The results of these measurements are presented in Fig. 2a where each plotted curve represents the average of 5 measurements at various positions along the designated areas. The inset in Fig. 2a gives the four areas where the spectroscopic data are acquired, that is to say in between two thin ribbons (red curve), at each edge of the thin ribbon (green and blue curves) and in the middle of the ribbon (black curve). There are two main information in the dI/dV spectrum in Fig. 2a about the density of state (DOS) peaks distribution in these thin ribbon structures . The first one shows that a DOS band energies lying between -3.0 V and - 2.4 V, have similar shapes and intensities. They correspond to a mix of DOS arising from the silicon subsurface and the observed layer[18]. The second one is more important and depicts several types of DOS peaks of various linewidths having their maximum voltage in the range 2.4 V – 1.5 V below the surface Fermi level energy. Looking with more details at the second group of DOS peaks, one can observe that the dI/dV curves are composed of two maximum at energies ~ -1.9 eV and ~ - 1.7 eV at the edges of the thin ribbon. However, only one DOS peak maximum at ~ 1.95 eV below $E_f$ is observed when the spectrum is acquired in between the thin ribbon (red curve) or inside of it (black curve). The very surprising difference in these two spectrums resides in their spectral linewidth energies. Indeed, when the dI/dV is acquired beside the periodic thin ribbons, the full width half maximum worth 490 meV while equal 210 meV inside the thin ribbon, which may reveals an electronic confinement of the localized density of state at this energy. Note that if one compares these results with the dI/dV signal acquired on a LRS (gray curve in Fig. 2a, structure reported in ref 7), our observations indicate that the thin ribbons exhibit a specific electronic structure that is clearly arising from their relatively small size and periodicity.

To further investigate the electronic properties of the observed periodic thin ribbons, it is necessary to estimate the insulating properties of such structure. Fig. 2b exhibits the tunnel current variation when acquired at various positions (see inset in Figs. 2b). On the WL (i.e. beside the periodic thin ribbons Fig. 1a), the surface gap energy is relatively small (~ 1.8 eV). This value can be considered as a landmark for comparison[8]. When the dI/dV curve is acquired in between the thin ribbons, the surface energy gap increases to ~ 2.8 eV (red curve in Fig. 2b). However, if the dI/dV signal is measured in the central part of the thin ribbon (black curve in Fig. 2b), or at its edges (green curve in Fig. 2b), the estimated surface gap energy is clearly larger than the previous measurements and worth 3.0 eV and 3.2 eV, respectively. Similar value of the surface energy gap is measured on LRS (~3.2 eV)[7] indicating that some parts of their structure may be similar. It is also interesting to notice that the energies of the valence band edges in Fig. 2b are very similar for each measured positions apart from the one at the wetting layer when located far from the periodic thin ribbons (see WL in Fig. 1a). Considering the SK process responsible for the growing of these structures, one can expect that the surface area located in between the thin periodic ribbons would be alike the one of the WL. However, the comparison of the conduction band edge of the red and black curves in Fig. 2b indicates that the insulating property of the surface area located in between the thin periodic ribbons is significantly increased compared to the one of the WL. This indicates that the electronic properties of the area located in between the thin ribbons may arise from a modified WL structure.

To clarify the presence or not of a DOS state enhancement in the center of the thin ribbons, we have measured the spatial distribution of the DOS on a surface area where several thin ribbons are formed (Fig. 3a). Simultaneously, the dI/dV signal at the resonance energy as observed in Fig. 2a (i.e. -1.95 eV) is acquired for each measured point of the DOS image. The resulting dI/dV map is shown in Fig. 3b. To help the comparison between Figs. 3a and 3b, red dotted lines are located on the central axis of the thin ribbons while green dotted lines indicate the common position situated in between the thin ribbons for the two panels. One can notice that, on the right part of the chosen area, one thin ribbon of stripes is slightly shifted (see the two white arrows) revealing that the periodicity between the thin ribbons can sometimes change on a given terrace. The first surprising observation in Fig. 3b is the presence of brighter lines of dI/dV signal delocalized along the central positions of the periodically spaced thin ribbons. However, while the dI/dV measurements acquired in between the thin ribbons reveal a DOS peak centered at the same resonant energy (Fig. 2), the dI/dV mapping shown in Fig. 3b mainly exhibits dark bands (i.e. along the green dotted lines, coherently to what is observed in Fig. 3a. Therefore, the enhancement of the DOS signal at the resonant energy 1.95 eV below the Fermi level is only observed at the central part of the thin ribbons (i.e. when the linewidth of the dI/dV peak worth 210 meV). Yet, one can also notice that, despite the slight shift in the ribbons periodicity mentioned previously (see white arrows in Fig. 3a), the observed enhancement can still be detected on the ensuing shifted thin ribbons. However, on the left side of the dI/dV mapping area where randomly arranged WL is covering the silicon surface, the dI/dV mapping exhibits no enhancement. These STM measurements confirm that the spatial distribution of DOS at this resonant energy is confined and delocalized along the central part of the thin ribbons. At this stage of our study, this resonant process seems to arise from the relatively small width of the observed ribbons structure.

To bring additional understanding of the observed periodic thin ribbons structure, we have built a theoretical model of a single thin ribbon unit cell and performed numerical simulations (i.e. annealing and relaxation) by using the density functional theory (DFT). Fig. 4a recalls a portion of the panel c in Fig. 1 in which one can observe various periodical or semi-periodical aspects in the thin ribbon structure (light brown and blue arrows). One of them (light brown arrows) indicates the alternative positions of bright spots located at the external edges of the thin ribbons, having a measured lateral period of ~14.9 Å. The dark blue arrows point out additional structures at the internal side of the thin ribbons (top panel in Fig. 4a) having a measured periodicity of ~7.5 Å. As a reminder, it is important to recall that the silicon dimer rows periodicity of the bare Si(100)-2x1 surface is 7.7 Å. A zoomed image of the unit cell of this structure is presented in the lower panel of Fig. 4a to be compared with the model structure of a thin ribbon unit cell shown in Fig. 4b. This structure is composed of a bare Si(100)-2x1 surface (blue atoms) on which has grown several WL unit cells having various orientations (see red dotted rectangles and their respective arrows). As previously recalled, the LRS structure, as detailed in a previous work[7], are formed by a superposition of WL unit cells, all oriented in the same direction, in between which are inserted additional $CaF_2$ molecules (see red dotted circles in Fig. 4b). The obtained structure forms stripes that expands perpendicularly to the Si dimer row as indicated by the green dotted rectangles. Here, the structure of the thin ribbon is formed via a specific orientation of these stripes as exposed in Fig. 4b. In particular, one can see that the central part of the thin ribbon is based of two head to tail unit cells. Each of them is followed by an additional unit cell having the same direction that forms the edges of the thin ribbon. Hence, the ensuing $CaF_2$ stripes of the thin ribbon follow the same orientation. One of the additional striking aspect of the thin ribbons (Fig. 4b) arises from the bright protrusions observed at the edges of the ribbon which position alternate from the left and right edges of the thin ribbon. We found that these features originate from the presence of additional $CaF_2$ molecules at the edges of the presented structure (see the light brown arrows reported in Fig. 4b).

To our knowledge, the formation of these periodically spaced superstructures on the Si(100) has never been reported yet. To bring more information on the process responsible of their formation, we have formed a second slab made of two parallel thin ribbons which structure recalls the unit cell of a single thin ribbon as described in Fig. 4b. The side and top view of this second structure are presented in Figs. 5a and 5c and related with the experimental data in Fig. 5b. Having a careful look in Fig. 5b reveals more details at the surface area located in between the thin ribbons. Indeed, one can observe that this surface region is not built similarly to the WL[8,9] (see for comparison at the left side in Fig. 1a). In particular, the STM topography of the surface area that separates the two thin ribbons exhibits a central continuous gray line with scattered dark spots. At this stage of our understanding, we found that this ensuing surface area is mainly composed of a band of three incomplete WL unit cells placed in a symmetrical way so that fluorine atoms passivate a silicon dimer line at equal distance of the two consecutive thin ribbons (Figs. 5a and 5b). Actually, by comparing Figs. 5a and 5c with Fig. 5b, one can observe that this second structure allows reproducing the dark grey line centered in between the two thin ribbons observed in Fig. 5b (follow the dotted green line) as well as the scattered dark spots appearing at each side of this line (Fig. 5d). Yet, one cannot completely exclude that some few missing silicon dimer arising from the initial stage of the WL layer formation may be formed in this area.

The proposed second structure of the double thin ribbon model as shown in Figs. 5a and 5c accurately describe our experimental finding and in particular, the alternated white spots observed at their edges. Indeed, the calculated local density of states (LDOS) of this structure shown in Fig. 5d for an integrated energy window of 2.5 eV below the Fermi energy can mimic the experimental measurement reported in Fig. 5b with a relatively good precision. It is important to notice that for the model of the two thin ribbons presented in Fig. 5, we have chosen to reproduce one of the main trends of our results, i.e. when the additional $CaF_2$ molecules are alternatively located at the two side-by-side edges of the thin ribbons. While it is possible to observe a conformation where the bright features at the edges of the thin ribbons are facing each other, their modelization would not change drastically our findings.

One major aspect of the formation of the periodically thin semi-insulating ribbons can be found in Fig. 1a where they arise at the edge of a silicon step edge. Since the formation of these periodic structures occurs when the substrate temperature is slightly higher than the formation of LRS (see section 2.1), one can suggest that the formation of the observed thin ribbons is related to both processes: (i) the epitaxy of the $CaF_2$ molecules on the surface and (ii) the Si(100) 2x1 surface reconstruction itself. Indeed, at the selected substrate temperature (~ 740 °C) the diffusion of silicon adatoms involved in the formation of the Si(100)-2x1 reconstruction is still very high, in particular along the [110] direction i.e. along the silicon dimer rows[23]. This temperature range involves especially the formation of the silicon step edges[24]. Since we have shown that the first stage formation of the wetting layer formation involves the etching of silicon dimers[7,8,9], the dynamics of the formation of the observed thin ribbons can clearly be influenced by a slight increase in substrate temperature.

From our observations, the previously reported LRS structure is observed when the silicon surface is heated at slightly lower temperature (710 °C) compared to the one allowing the formation of the periodically spaced thin ribbons of stripes (740 °C). This relatively small variation of temperature already involves observable differences in the epitaxial surface structure, especially on the WL formation, while the CaF$_2$ coverage is kept constant in both cases. Indeed, as observed in the STM topography in Fig. 1a, we can see that the silicon surface is fully covered by a WL interface. In contrast, remaining small patches of bare silicon surface are surrounding the LRS[7]. This change indicates that the growth switches from a Volmer-Weber (VM) to a Stranski-Krastanov (SK) regime when the temperature slightly increases as mentioned above. Since the misfit between the bulk Si and CaF$_2$ is rather small (i.e. 0.021 Å) the observed growth regime change is thus mainly due to a surface energy alteration[25, 26]. In our particular case, the surface free energy can significantly vary due to the temperature change, as explained previously because of the increasing diffusion of Si adatom when the surface temperature reaches 740 °C. The surface energy of the Si(100)-2x1 can also be affected by the etching process via the CaF$_2$ dissociation that occurs at the first stage of the epitaxy[8]. The initial value of the free energies of the two components worth $\gamma_{s[Si(100)-2x1]}$ = 2.7 meV/Å and $\gamma_{0[CaF2(100)]}$ = 5.7 meV/Å where $\gamma_s$ and $\gamma_0$ are the surface energy at the silicon substrate-vacuum interface and the overlayer-vacuum interface, respectively[27]. Usually, when $\gamma_0 > \gamma_s$, the VW regime is predominant, which is coherent with our observations when the silicon substrate is kept at a temperature below 710° C with the growth of LRS structures. However, in the case of a periodic thin ribbon formation, the slight increase in temperature will promote the formation of a full WL, which will completely degrade the initial Si(100)-2x1 surface reconstruction, resulting in an additional increase of the silicon surface energy $\gamma_s$ to lead to the SK regime[28].

Nevertheless, the two previous parameters cannot alone be responsible for the observed periodicity measured between two thin ribbons of stripes (i.e. 55 Å) over areas extending up to 15-20 ribbons. A careful look at the profile in Fig. 1d indicates that the thin ribbon width (i.e. ~26 Å) is about the half of the averaged periodicity between two thin ribbons. Since a thin ribbon is, itself, composed of four WL unit cells, this observation indicates that the distance separating two thin ribbons is also related to the size of a WL unit cell. Finally, one can observe that the surface separating two thin ribbons remains darker upon STM topography, indicating that several substitutional Ca atoms are probably missing in these areas as discussed above. Therefore, one can assume that the additional $CaF_2$ molecules that aggregate along the edges of the thin ribbons may arise from the surface area formed by the incomplete WL separating two thin ribbons.

The specific orientation of the $CaF_2$ stripes in the thin ribbon structures, as described in Fig. 4b, may also arise from the influence of the dipole effects on the polar Ca, F and $CaF_2$ components involved in the thin ribbons. To do so, we have plotted a three-dimensional representation of the atomic forces applied on each Ca atom of a slab composed of two consecutive thin ribbons (Fig. 6a). The size of the considered slab is depicted by the red dotted rectangle shown in Fig. 5c. It is composed of 8 rows of 5 Ca atoms (see section 2.2 for details). Each Ca atom is numbered from 1 to 36 in Fig. 5c. For a better interpretation, the components of each atomic forces are projected over the three 2D planes XY, XZ and YZ. The Fig. 6b presents the projection of the atomic forces on the XY plane and shows the relatively low values of these forces in this plane compared to the z component shown in the other panels in Fig. 6. For guidance, the number of some Ca atoms are recalled. Interestingly, one can observe that the direction of the atomic forces applied on the substitutional Ca atoms of the slab (i.e. Ca N° 2, 16, 20 and 34, green circles) are in opposite direction inside the same thin ribbon (Ca N° 2 and 16) but also for Ca composing two consecutive thin ribbons (Ca N° 16 and 20). This is a first indication that the relative orientation of the WL unit cells, as depicted in Figs 4b and 5c, stabilizes the structure of the thin ribbon. As well, this configuration also stabilizes the interaction between thin ribbons. One can also notice that the Ca atoms forming the $CaF_2$ molecules along the stripes show only weak atomic force components in this plane (see the green dotted circles in Fig. 6b for the Ca N° 4, 18, 22, 35). Similarly, all the Ca atoms that compose the internal part of the thin ribbon are subject to weak atomic forces.

Surprisingly, the atomic forces on some of the Ca atoms exhibit components that are mainly perpendicular to the silicon surface (Figs. 6c and 6d). These forces appear to be especially acting on the Ca atom numbers 3, 15, 21, 33 when the z-component is directed upwards. These Ca atoms implicate the $CaF_2$ molecules that agglomerate at the edges of the thin ribbons. Yet, the z-components of the atomic forces directing downwards the surface act on the Ca atom numbers 4, 14, 22 and 32. Here, the involved Ca atoms correspond to some of the $CaF_2$ molecules that intercalate between the WL unit cells, along the stripes. Hence, it appears that the additional free $CaF_2$ molecules that decorate the edges of the thin ribbons stabilize the structure by compensating the forces in the opposite direction. The panel d in Fig. 6 mainly corroborates these trends. In particular, looking at the relative heights of the various Ca atoms shown in the insets in Fig. 6c and 6d helps to clarify the positions of the discussed Ca atoms.

It would be extremely interesting to further study the influence of various parameters such as the precise substrate temperature, the $CaF_2$ flux or the coverage rate on the growth of the thin insulating ribbons. Moreover, the electronic confinement observed in these thin ribbons of stripe could reveal specific physical effects such as anti-corrugating state resonances that could be enhanced via the use of such structures[18]. The study of these effects may surely be the purpose of future investigations, but so far, not in the scope of the present article.

## 4 CONCLUSION

In conclusion, we have observed the growth of periodically spaced thin ribbons when the bare Si(100)-2x1 surface is exposed to $CaF_2$ molecules at a specific substrate temperature range (~ 740 °C). These structures, investigated and analyzed via STM techniques and reproduced by DFT methods, exhibit a periodicity of about half their width over distances that can exceed 80 nm. The observed thin ribbons show semi-insulating properties similar to LSR previously studied (i.e. ~3.2 eV). One of their particularities is that the thin ribbons exhibit a resonant state at ~ 1.95 eV below the surface Fermi energy with a relatively low linewidth of 210 meV. Such resonant DOS peak could be exploited via the opening of specific planar transport channels in molecular adsorbates by adding contacting pads nearby these structures[15]. Considering the ensemble of their properties, the observed periodic semi-insulating thin ribbons may be attractive in various scientific communities having fundamental interests in condensed matter physics, surface chemistry or molecular electronics as well as, for more applied purposes, those dealing with microelectronic including FINFET transistors that are nowdays reaching sizes that are comparable to the observed ribbons width. Our finding thus paves the way to a very interesting novel method that will allow to precisely localizing not only insulating structures but also any islands that uses epitaxy techniques in the VM or SK growth regimes.

**ASSOCIATED CONTENT**

There is no associated content for this article.

**AUTHORS INFORMATION**

**Corresponding author:**

Dr. Damien RIEDEL, Email: damien.riedel@u-psud.fr, ORCID N° 0000-0002-7656-5409

Author:

Dr. Eric Duverger: ORCID N° 0000-0002-7777-8561

**Author contributions:**

DR performed the experimental measurements and data analysis and mainly wrote the article. ED performed the DFT simulations on the large atomic slab. DR participates as well to the DFT simulations discussions. All the authors contributed to the writing and proofreading of the manuscript.

**ACKNOWLEDGEMENTS**

All the authors wish to thank the French National Agency for Research (ANR) for their financial support of the CHACRA project under the contract number ANR-18-CE30-0004-01. This work was supported by the Mesocentre of Université de Franche Comté that granted its access for DFT simulations. DR would also like to thank the Mésocentre LUmière Matière (MésoLUM) at ISMO for granting the access to their computational resources.

**FIGURES CAPTIONS**

**Figure 1:** (a) 93.5 x 93.5 nm² STM topography (-2.4 V, 15 pA) showing a large-scale area of thin ribbons. (b) 46.8 x 46.8 nm² STM topography (-2.4 V, 16 pA) of a zoomed-in area indicated by the orange square in (a). (c) 110 x 110 Å² STM topography (-2.4 V, 16 pA) of two consecutive thin ribbons. The red dotted rectangle indicates the area on which lateral lines profiles (red line) have been measured and averaged. (d) Variation of the relative height acquires across two thin ribbons as shown in (d) as a function of the lateral distance.

**Figure 2:** Variation of the dI/dV signal as a function of the bias voltage applied on the sample surface at several locations on (blue, black, green dots) or beside (red dots) the thin ribbons structure. (b) Logarithmic scale variation of the tunnel current as a function of the bias voltage for the four indicated positions as shown in the inset in (b). The blue curve is the I(V) signal acquired on the WL area located far beside the thin ribbons as shown in the left part in Fig. 1a. The values indicated at each edge of the curves are in mV.

**Figure 3:** (a) 357 x 176 Å² STM topography (-2.3 V, 16 pA) off six parallel thin ribbons with a small WL on the left. (b) dI/dV map of the same area as shown in (a) at the bias -1.95V (see methods for details). The green or red dotted lines are landmarks to locate the positions of the thin ribbons.

**Figure 4:** (a) 56.5 x 34 Å² STM topography (-2.4 V, 16 pA) centered on a small part of a thin ribbon. Blue and light brown arrows indicate the location of the internal or external structure of the ribbon, respectively. The red dotted rectangle shows the zoomed portion of the image presented below. (b) ball and stick representation of the model structure reproducing the thin ribbon shown in the lower panel of (a). A detailed WL unit-cell is recall below the figure and its size are reported on the slab via the red dotted rectangles. The red horizontal arrows designate the relative orientation of the WL unit-cell. The green dotted rectangle indicates the stripe structure. Two superimposed red broken lines to follow the relative positions of the $CaF_2$ molecules located at the edges of the thin ribbon. The cyan, blue and gray balls are Ca, Si and F atoms, respectively.

**Figure 5:** (a) Ball-and-stick representation (side view) of a structure composed of ten silicon layers and two thin ribbons. (b) 102 x 33 Å² STM topography (-2.45 V, 15 pA) of two consecutive thin ribbons. (c) Balls and sticks representation (top view) of the same slab as in (a). (d) Calculated local density of states image of the slab presented in (a) and (b) at an energy of 2.5 eV below the Fermi level. The cyan, blue and light grey balls are Ca, Si and F atoms, respectively in the panel (a) and (b). The light brown arrows indicate the relative position of the side additional $CaF_2$ molecules relatively to the bright protrusions observed on STM topographies.

**Figure 6 (two columns):** (a) Three dimensional vectorial representation of the calculated atomic forces (eV/Å) applied on the Ca atoms of the slab as defined in figure 5c of a double thin ribbon. (b) to (d) two dimensional projection of the calculated atomic forces vectors in (a) on the XY, XZ and YZ planes, respectively. The insets in the panels (c) and (d) recall a simplified atomic slab of the two thin ribbons model as defined in figure 5c.

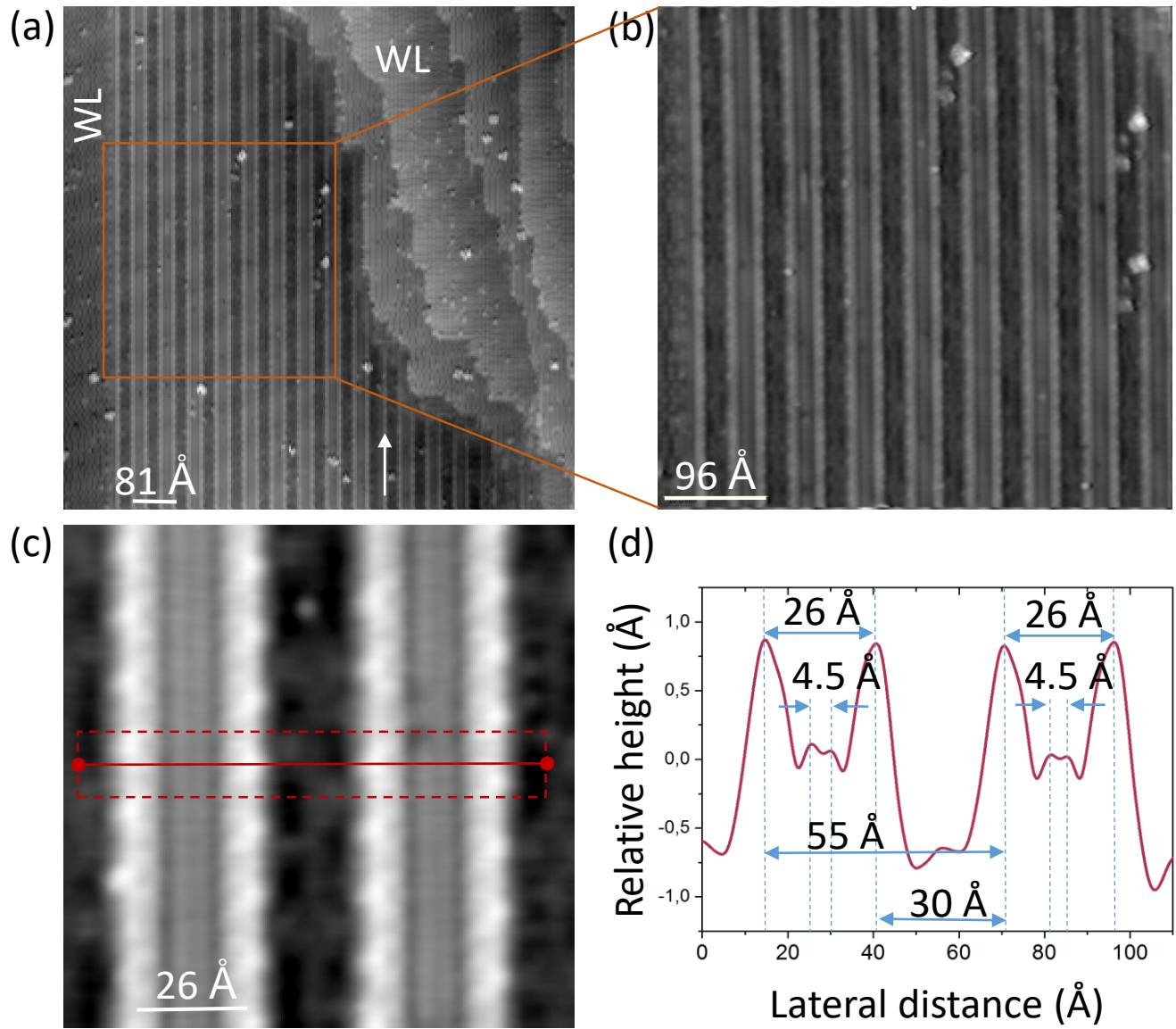

Riedel et al. Figure 1

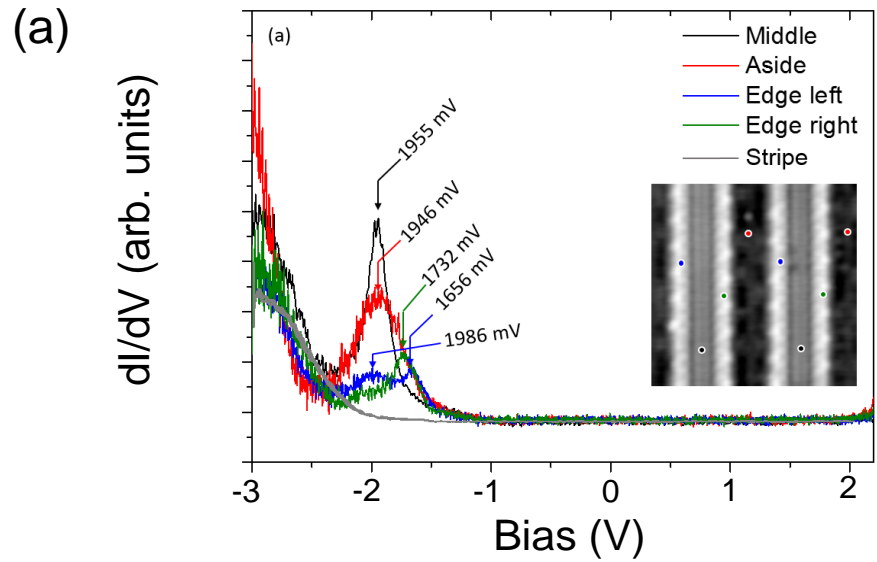

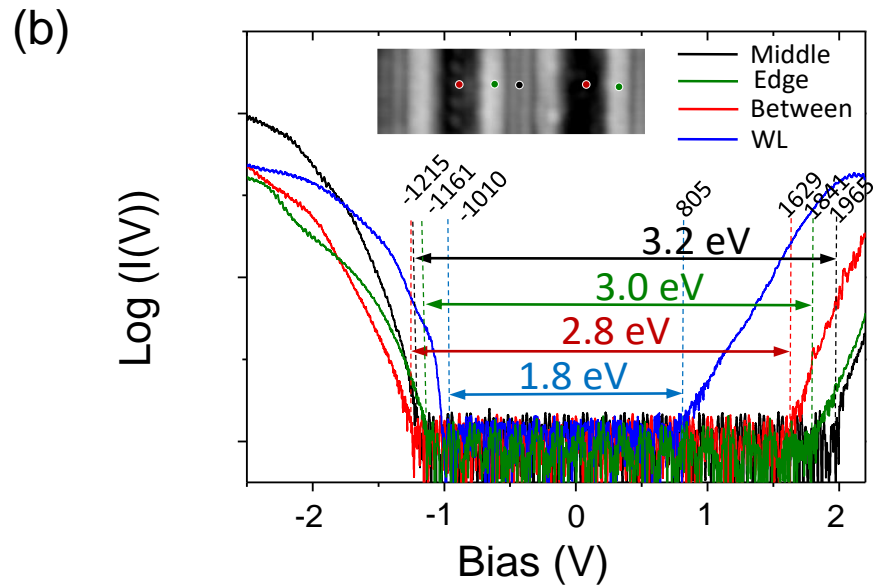

Duverger et al. Figure 2

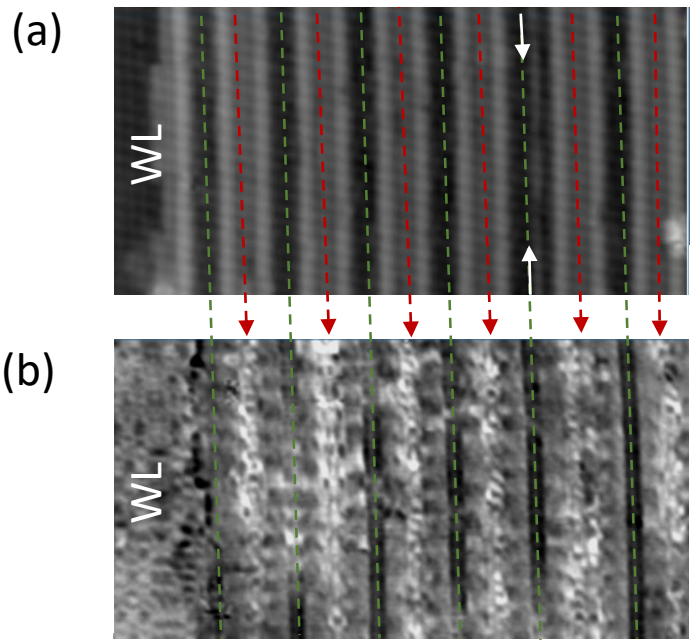

WL

(a)

WL

(b)

Duverger et al. Figure 3

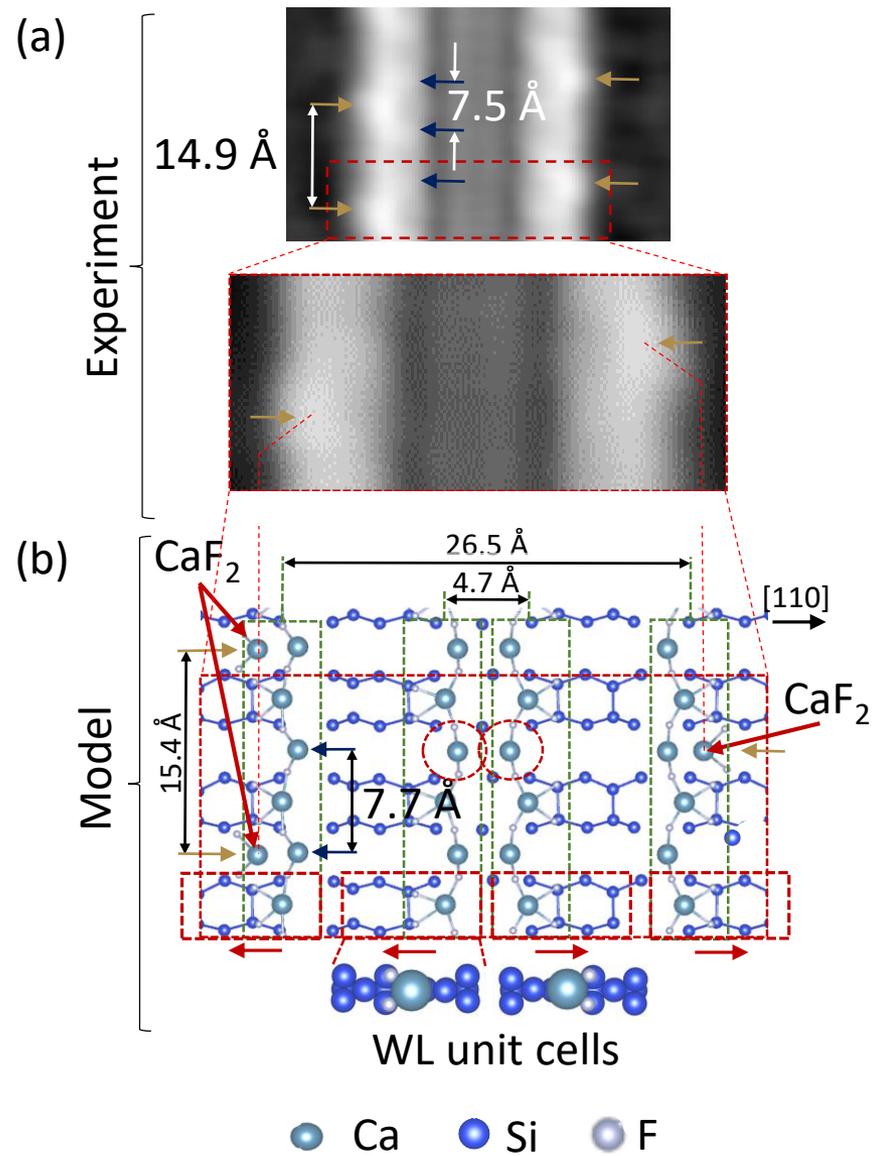

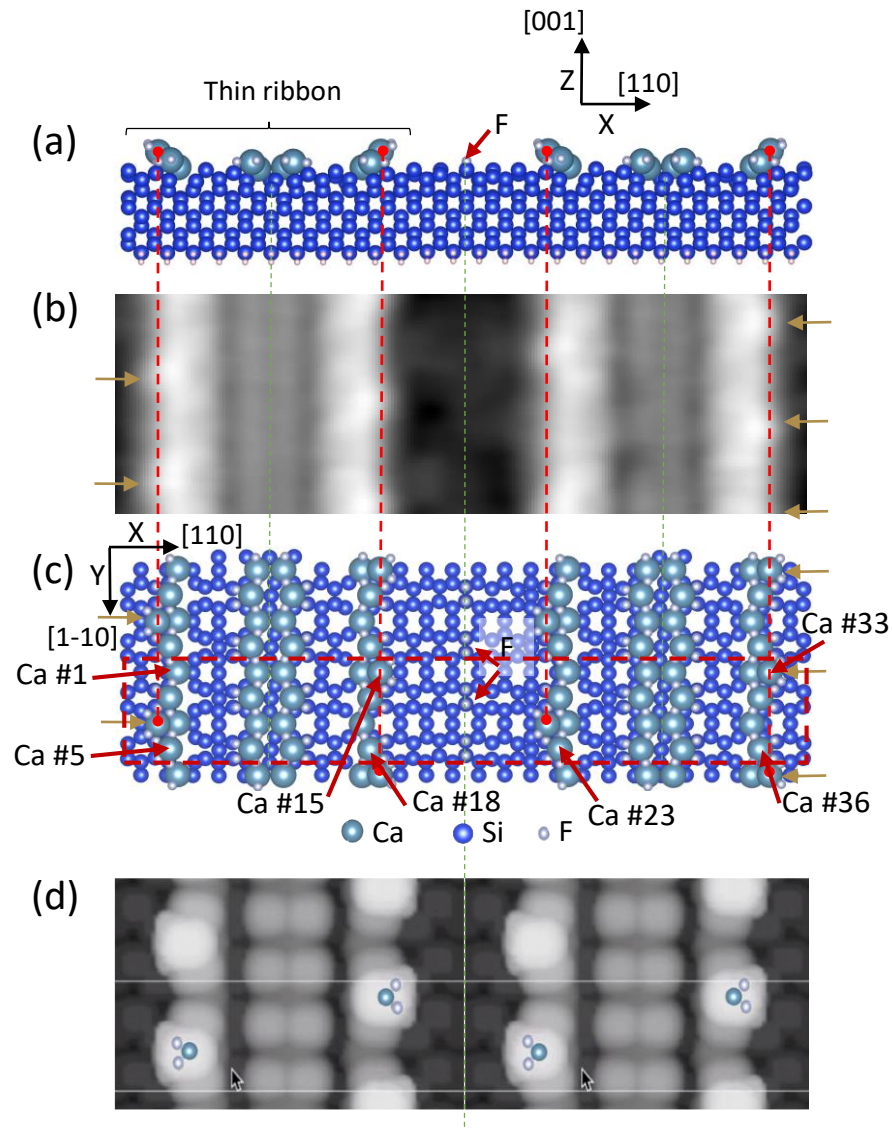

Duverger et al. Figure 5

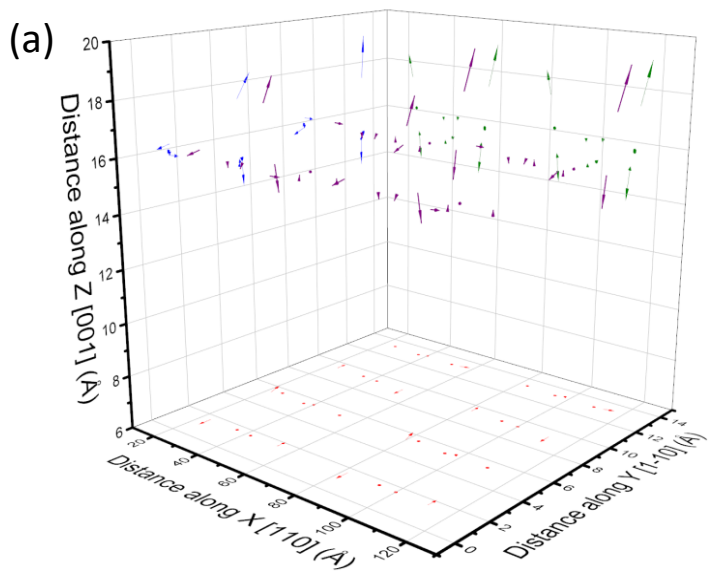

(a)

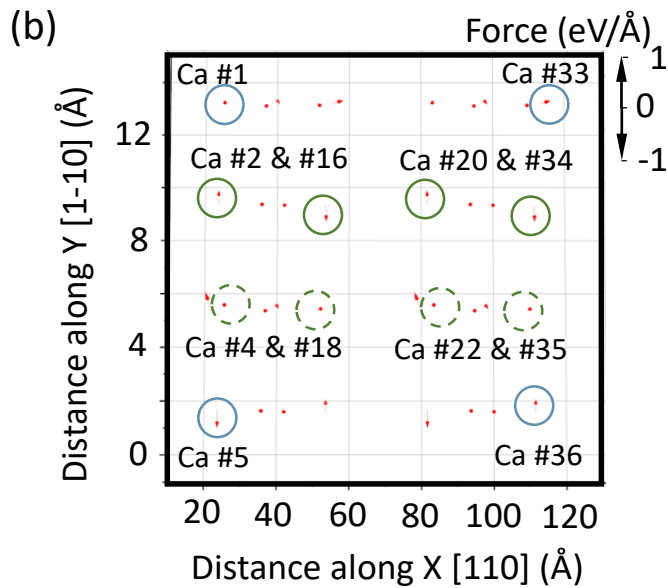

(b)

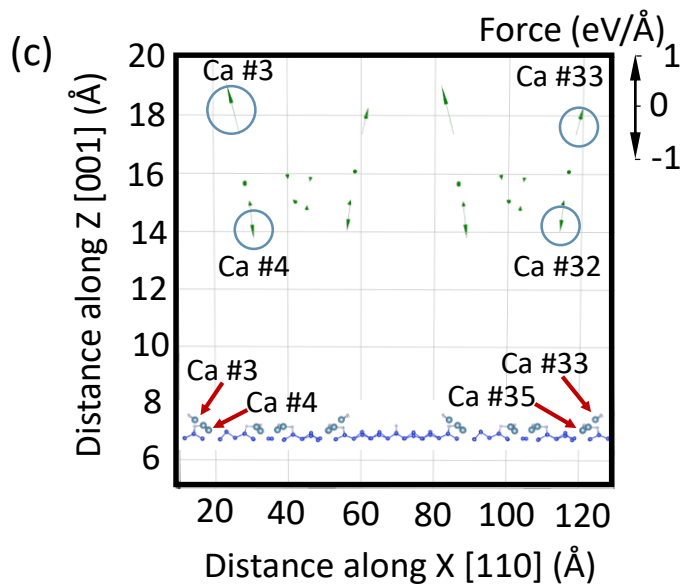

(c)

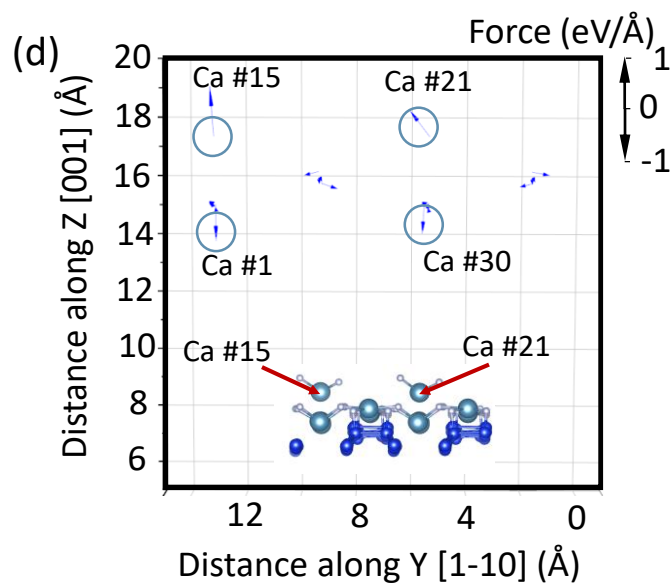

(d)

Duverger et al. Figure 6